\documentclass[11pt]{elsarticle}

\usepackage{setspace}
\usepackage{amsfonts}
\usepackage{amssymb}
\usepackage{algpseudocode}
\usepackage{mathrsfs}
\usepackage{multirow}
\usepackage{tikz}
\usepackage{setspace}
\usepackage{etoolbox}   
\usepackage{lipsum}
\usepackage{booktabs}
\usepackage{graphicx,wrapfig}
\usepackage{mathtools}
\usepackage{comment}
\usepackage[hyphens]{url}
\usepackage[hyphenbreaks]{breakurl}

\AtBeginEnvironment{tabular}{\doublespacing}

\usetikzlibrary{arrows,automata,backgrounds,fit,patterns,petri,shadows,shapes, positioning}

\usepackage[top=3.5cm,bottom=3.5cm,left=3.5cm,right=3.5cm]{geometry}

\usepackage{amsthm}

\usepackage{lineno}
\usepackage[hidelinks]{hyperref}
\modulolinenumbers[5]

\usepackage{amssymb}
\usepackage{mathtools}
\usepackage{mathrsfs}  
\usepackage{float}
\usepackage{graphicx}  

\usepackage{amsmath}
\usepackage{graphicx,psfrag,epsf}
\usepackage{enumerate}
\usepackage{natbib}
\usepackage{comment}
\usepackage[official]{eurosym}
\usepackage{eurosym}
\usepackage{url} 
\usepackage{siunitx}
\usepackage{bbm}

\usepackage{url}            
\usepackage{booktabs}       
\usepackage{subcaption}
\usepackage{amsfonts}       
\usepackage{nicefrac}       
\usepackage{microtype}      

\usepackage{amsmath}
\usepackage{mathabx}
\usepackage{amsopn}

\usepackage{algpseudocode}
\usepackage{algorithm}

\algrenewcommand\algorithmicindent{0.5em}%

\makeatletter
\renewcommand{\Function}[2]{%
  \csname ALG@cmd@\ALG@L @Function\endcsname{#1}{#2}%
  \def\jayden@currentfunction{#1}%
}

\newcommand{\funclabel}[1]{%
  \@bsphack
  \protected@write\@auxout{}{%
    \string\newlabel{#1}{{\jayden@currentfunction}{\thepage}}%
  }%
  \@esphack
}

\usepackage{color}
\usepackage{caption}

\usepackage{float}
\floatstyle{plaintop}
\restylefloat{table}

\usepackage{amsthm}
\usepackage{bm}

\usepackage{color,soul}
\usepackage[usestackEOL]{stackengine}[2013-09-11]

\setstackgap{L}{1.4\baselineskip}
\usepackage{tikz}
\usetikzlibrary{arrows,automata,backgrounds,fit,patterns,petri,shadows,shapes, positioning}
\pgfdeclarepatternformonly{stripes}
{\pgfpointorigin}{\pgfpoint{0.25cm}{0.25cm}}
{\pgfpoint{0.25cm}{0.25cm}}
{
    \pgfpathmoveto{\pgfpoint{0cm}{0cm}}
    \pgfpathlineto{\pgfpoint{0.25cm}{0.25cm}}
    \pgfpathlineto{\pgfpoint{0.25cm}{0.12cm}}
    \pgfpathlineto{\pgfpoint{0.12cm}{0cm}}
    \pgfpathclose%
    \pgfusepath{fill}
    \pgfpathmoveto{\pgfpoint{0cm}{0.12cm}}
    \pgfpathlineto{\pgfpoint{0cm}{0.25cm}}
    \pgfpathlineto{\pgfpoint{0.12cm}{0.25cm}}
    \pgfpathclose%
    \pgfusepath{fill}
}

\usepackage{tabularx}
\usepackage{booktabs}

\begin{document}

\begin{frontmatter}

\title{On Large Language Models in National Security Applications}


\author[my1address]{William N. Caballero}
\author[my1address]{Phillip R. Jenkins}

\address[my1address]{Department of Operational Sciences, Air Force Institute of Technology, WPAFB, OH 45433}

\begin{abstract}
The overwhelming success of GPT-4 in early 2023 highlighted the transformative potential of large language models (LLMs) across various sectors, including national security. This article explores the implications of LLM integration within national security contexts, analyzing their potential to revolutionize information processing, decision-making, and operational efficiency. Whereas LLMs offer substantial benefits, such as automating tasks and enhancing data analysis, they also pose significant risks, including hallucinations, data privacy concerns, and vulnerability to adversarial attacks. Through their coupling with decision-theoretic principles and Bayesian reasoning, LLMs can significantly improve decision-making processes within national security organizations. Namely, LLMs can facilitate the transition from data to actionable decisions, enabling decision-makers to quickly receive and distill available information with less manpower. Current applications within the US Department of Defense and beyond are explored, e.g., the USAF’s use of LLMs for wargaming and automatic summarization, that illustrate their potential to streamline operations and support decision-making. However, these applications necessitate rigorous safeguards to ensure accuracy and reliability. The broader implications of LLM integration extend to strategic planning, international relations, and the broader geopolitical landscape, with adversarial nations leveraging LLMs for disinformation and cyber operations, emphasizing the need for robust countermeasures. Despite exhibiting ``sparks" of artificial general intelligence, LLMs are best suited for supporting roles rather than leading strategic decisions. Their use in training and wargaming can provide valuable insights and personalized learning experiences for military personnel, thereby improving operational readiness.
\end{abstract}

\begin{keyword}
Large language models \sep applied statistics \sep artificial intelligence \sep national security \sep strategic planning
\end{keyword}

\end{frontmatter}

\section{Introduction}

The profound success of GPT-4 in early 2023 \citep{roose2023gpt} demonstrated the revolutionary nature of large language models (LLMs) to a global audience. Their capabilities were immediately recognized as a potentially disruptive force in numerous industries, e.g., healthcare \citep{jindal2024large}, education \citep{manning2024original}, and finance \citep{macrae2024large}. However, as with other artificial intelligence (AI) models, LLMs are not limited in scope to civilian applications; their capabilities may also transform national security operations. Indeed, as Richard Moore of MI6 highlighted, AI, including LLMs, is already a critical factor in current threat environments, acting as a force multiplier for many tools and practices \citep{Sciutto2023}.

The nature and degree of this transformation have yet to be determined, but speculation abounds. For example, using AI in defense applications is a disquieting topic for some observers, conjuring pop-culture images of \textit{Skynet} \citep{MarkoffRosenberg2016}. However, numerous applications lacking autonomous weapon employment can be drastically reshaped through AI integration. Such applications overlap with civilian activities (e.g., natural language processing tasks,  computer-vision implementation, and decision-analytic problems), domains where LLMs are being rapidly commercialized. Thus, opinions on using LLMs in defense applications run the gamut; proponents contend that they may be used to create a virtual Clausewitz, whereas dissidents believe their use in any defense setting would result in dangerous hallucinations. Similar contentions arise when considering using LLMs outside of military activity and within the use of other national security instruments, e.g., diplomatic, information, and economic measures \citep{JointChiefsJP30}. 

From the perspective of a statistician or data scientist, applying LLMs to national security problems will feel quite familiar despite the domain's derived idiosyncrasies. Nearly every civilian activity has a national security analog, implying that LLMs developed for a particular civilian task may be extended to a national security setting after domain-specific modification. Military physicians may receive decision support from LLMs, provided these systems have been tuned to the realities of combat. Likewise, LLMs may support the decision-making of military commanders via automatic summarization, sentiment analysis, and topic modeling, provided they have been trained on domain-specific vocabulary, acronyms, and jargon. However, the national security setting modifies the relative importance of select LLM design challenges. For example, even if an LLM is only used for automatic summarization during armed conflict, the effect of hallucinations may be catastrophic if these summaries inform senior-level decision-making. On another extreme, if an LLM is used to provide strategy or planning recommendations, then model interpretability and output transparency are paramount to ensuring effective human oversight. The same applies to understanding emergent LLM abilities in national security applications as well \citep{wei2022emergent, schaeffer2024emergent}. Moreover, when an LLM has access to classified information, the importance of data privacy and the relevance of prompt-injection attacks \citep[e.g., see][]{liu2024automatic} increases dramatically; if the underlying information is not appropriately safeguarded, security incidents (e.g., data spillage) that endanger national security personnel and the civilian population pose a significant risk. The national security setting is high stakes and leaves little margin for error in LLM applications.

Therefore, within this manuscript, we explore the ramifications of LLMs on national security applications viewed in terms of the strategic competition continuum set forth by the \citet{JointChiefs2018}. This framework does not view international conflict in binary terms (e.g., war and its absence) but interprets interactions between rivals continuously varying from peaceful cooperation to armed conflict across multiple domains. We synthesize varying perspectives on the use of LLMs for national security applications, examine the current usage of LLMs in defense settings, explore their efficacious use in defense applications, and forecast their more general effect on national security writ large. Additionally, whereas LLMs are the focus of this manuscript, emphasis is also placed on the interface of LLMs with alternative probabilistic, statistical, and machine learning (ML) methods to determine how this interplay may affect national security operations.

The remainder of this manuscript is structured as follows. Section \ref{sec2} inspects the contemporary application of LLMs by the national security establishment and reviews its relation to scholarly research on the topic. In this manner, we tangibly explore the reception of LLMs among a subset of national security organizations and investigate how they relate to the views of other national security scholars. Based upon this information, Section \ref{sec3} provides our views on how LLMs can be best integrated into defense applications and discusses their broader repercussions across the other instruments of national power. Section \ref{sec4} provides concluding remarks and posits avenues of future inquiry. 

\section{A Rapidly Evolving and Uncertain Landscape} \label{sec2}

Whereas the impressiveness of LLMs is self-apparent, at the time of this writing, their utility in application is less certain. This is typified by Google's difficulties incorporating Gemini, a multi-modal LLM, into its search engine \citep{Dobuski2024} and recent surveys on generative AI utilization \citep{ZaoSanders2024}. Nevertheless, the LLM landscape is rapidly evolving, inducing numerous stakeholders to experiment with associated transformations to national security operations. This section reviews such experimentation and their relation to the scholarly literature. 

\subsection{Recent LLM Initiatives and Applications by National Security Organizations}

From the perspective of the US Department of Defense (DoD), the emergence of high-functioning LLMs was perfectly timed for innovation. After nearly two decades focused on counterterrorism operations, the 2018 National Defense Strategy (NDS) refocused the DoD on inter-state strategic competition. This strategy recognized that the military advantage enjoyed by the US after the Cold War had eroded \citep{mattis2018summary} and, as one means to rectify this realty, stated that the DoD would ``invest broadly in military application of autonomy, artificial intelligence [AI], and machine learning [ML], including rapid application of commercial breakthroughs.'' Six years after its publication, the acquisition practices exposed by the 2018 NDS, which are also expressed in the 2022 NDS, have taken root, and their effects are apparent in the DoD's approach to LLMs.

In August 2023, Task Force (TF) Lima was created under the DoD Chief Data and AI Office (CDAO) with the express purpose of identifying low-risk applications for generative AI and LLMs \citep{Vincent2023}.  Such a focus on low-risk areas coincides with recent strategic guidance from \citet{biden2023} and the \citet{DoD_RAI_Strategy} on responsible AI use; in the view of Deputy Secretary of Defense Kathleen Hicks, ``most commercially available systems enabled by [LLMs] aren’t yet technically mature enough to comply with our DoD ethical AI principle.'' At the time of this writing, TF Lima's work continues; however, it has highlighted the DoD's use of LLMs to streamline staff operations within its internal bureaucracy via, e.g., the automatic summarization of doctrine, instructions, and policy manuals. TF Lima is also endeavoring to identify how LLMs affect classification guidance, especially given the effects of classification by combination, i.e., a national security particularity whereby specified unclassified information becomes classified upon combination \citep{Vincent2024}. In line with such concerns, TF Lima is also constructing a ``virtual sandbox" through which military personnel can experiment with generative AI tools \citep{DefenseScoop2024}; suitably, TF Lima has its own ChatGPT instance that can be used to discuss its activities \citep{TFLIMA_2024}.

Despite the ongoing development of a unified experimentation platform, the pace of LLM development has necessitated that other DoD agencies work in parallel with TF Lima. The United States Air Force (USAF) has been particularly active, and LLMs featured chief among the topics discussed at the USAF's 2024 Data, Analytics, and AI Forum. Multiple USAF organizations have developed proof of concepts whereby LLMs are leveraged to expedite myriad coding and administrative tasks. For example, Air Mobility Command has leveraged LLMs to generate campaign simulations based on user-defined, text-based narratives. Moreover, US Air Forces Central, among other USAF organizations, use LLMs to expedite routine maintenance of their endemic software tools. More recently, at the Bravo 11 Hackathon, a US Pacific Air Forces team spearheaded an ambitious use case whereby LLMs automatically summarize incoming mission reports to rapidly distill insights for commanders \citep{Leidos2024, ross2024}. In so doing, the team illustrated how an LLM  can readily handle a task that, when performed manually, requires multiple operators and weeks of man-hours. Similar time savings were shown at the Air Force Test Center's Data Hackathon when an LLM pipeline was created to automatically generate flight test documents, e.g., plans and reports \citep{conner2024}; the prototypes were based on the relatively small MPT-7b, MPT-30b, Falcon-7b, and Falcon-40b models augmented with a customized retrieval augmented generation (RAG) architecture. 

Additionally, Air University's innovation center (AUiX) has been developing a GPT framework for wargaming called the Comprehensive Heuristic Utility for Combat Knowledge (CHUCK), which is currently in a beta stage but shows exciting potential for the future. This initiative is part of a broader collaboration with Stanford University's Hoover Institution and the MIT Artificial Intelligence Accelerator (MIT/AIA), aimed at advancing wargaming techniques and exploring LLM's impact on crisis decision-making. These efforts align with the Air Force Futures office's AI initiatives, exploring how AI can enhance wargaming by running thousands of iterations to optimize strategies and decision-making \citep{AirForceAIWargaming2024}. Such initiatives aim to transform traditional wargaming by incorporating  AI capabilities to improve strategic analysis and operational planning. In conjunction with these wargaming efforts, the Air Force Research Laboratory also developed NIPRGPT, an LLM recently cleared and deployed by the USAF for installation on designated unclassified systems \citep{AirForceGenerativeAI2024}. 

Alternatively, the USAF's most innovative use cases derive from its academic institutions. Cadets in the Data Science program at the USAF Academy constructed an LLM-based prototype to modernize the user interface in the USAF's Envision platform \citep{AirSpaceForces2023}. The prototype is a ``no-code" solution allowing users to submit statistical queries via text-based narratives and receive outputs with corresponding explanations. For example, a user may request a graphical summary of pilot training graduation rates by commissioning source, and the tool may output a box plot with an associated narrative on how to interpret the figure. The USAF Academy has also used LLMs for assessments within its statistical courses. In the Spring 2024 semester, LLMs were used in its applied statistical modeling class (i.e., MATH 378) to conduct ``oral boards" instead of examinations; that is, students were asked to interface with an LLM on a specified set of topics, and the associated transcripts were used to assess student knowledge. LLMs are being further integrated into USAF Academy instruction via the development of a virtual teaching assistant. Notably, seniors in the Data Science program created the QuantaIQ prototype that uses LLMs to generate assignments, conduct assessments, and interface with students via question-and-answer sessions. 
 
Other branches of the United States military are also exploring LLMs, though their approaches and applications vary. The United States Army is experimenting with generative AI in military video games to improve battle planning by using LLMs, including OpenAI’s GPT-4 Turbo and GPT-4 Vision models, to provide information on battlefield terrain and details on friendly and enemy forces \citep{ScienceDesk2024}. Additionally, the Army is developing new policy guidance to guide the department's use of generative AI to streamline operations while addressing security concerns \citep{Harper2024}. In collaboration with Scale AI, the Marine Corps created an LLM named Hermes to assist in military planning by synthesizing data, generating hypotheses, and refining courses of action \citep{JensenTadross2023}. At Marine Corps University, experts are incorporating LLMs into simulations and wargames to test whether they improve analytical products and ease of use for military students \citep{Vincent2023_marine}. These experiments demonstrated that LLMs can enhance the planning process and provide valuable insights but also require human oversight to manage current limitations, such as hallucinations and structural biases, to ensure accuracy. 

In contrast, the United States Navy has adopted a more cautious approach to using LLMs. According to a recent memo published by the Navy’s acting Chief Information Officer, ``the use of proprietary or sensitive information poses a unique security risk and has the potential to lead to data compromise when employed by commercial generative AI models" \citep{Rathbun2023}. The Navy's policy advises against using commercial LLMs for operational purposes until security requirements are fully investigated and approved \citep{Errick2023}. This cautious stance is reflected in efforts to secure access to LLM technology through Jupiter, the Navy's enterprise data and analytics platform, to ensure safe employment \citep{OConnell2023}. Similarly, the United States Space Force implemented a temporary ban on using generative AI and LLM tools for official purposes in October 2023 due to concerns about safeguarding sensitive data \citep{Harpley2023}. As of the time of this writing, the ban remains in place as the service continues to evaluate the best path forward for securely integrating generative AI capabilities into its mission \citep{Harpley2024}.

Beyond the DoD, other US national security agencies are also exploring LLM applications. The Central Intelligence Agency (CIA) began exploring generative AI and LLM applications more than three years before the widespread popularity of ChatGPT. For example, generative AI was leveraged in a 2019 CIA operation called Sable Spear to help identify entities involved in illicit Chinese fentanyl trafficking \citep{Bajak2024}. The CIA has since used generative AI to summarize evidence for potential criminal cases, predict geopolitical events such as Russia's invasion of Ukraine, and track North Korean missile launches and Chinese space operations \citep{Bajak2024}. In fact, Osiris, a generative AI tool developed by the CIA, is currently employed by thousands of analysts across all eighteen U.S. intelligence agencies. Osiris operates on open-source data to generate annotated summaries and provide detailed responses to analyst queries \citep{Bajak2024}. The CIA continues to explore LLM-incorporation in their mission sets and recently adopted Microsoft's generative AI model to analyze vast amounts of sensitive data within an air-gapped, cloud-based environment to enhance data security and accelerate the analysis process \citep{Suciu2024}. Other agencies, including the Defense Advanced Research Projects Agency (DARPA), which uses LLMs to detect and fix critical software vulnerabilities, and MITRE, which collaborates with NATO to enhance AI security, are also exploring the effects of LLMs in their application areas \citep{Vergun_2024, Trifiletti2023}.


Moreover, allied countries are actively investigating LLMs to enhance military capabilities across various domains as well. For example, the 2024 TIDE Hackathon, co-hosted by NATO Allied Command Transformation (ACT) and the Dutch Ministry of Defence, included an LLM wargaming challenge \citep{NATO2023}. This challenge focused on using LLMs to improve military wargaming by creating dynamic scenarios and providing real-time feedback to enhance interoperability, decision-making, and strategy development among allied forces. LLM capabilities have garnered attention from other NATO entities aside from ACT as well. Namely, the NATO Communications and Information Agency (NCI Agency) is developing an AI cognitive agent to automate routine IT-support tasks \citep{Harper2023}. Similarly, Hadean, a British deep tech start-up company, secured a contract with the Defence and Security Accelerator (DASA) to develop an LLM for the British Army's virtual training space \citep{Hadean2024}. This initiative seeks to create a dynamic virtual environment with realistic human terrain and social media simulation, providing real-time feedback, generating complex scenarios, and assisting in after-action reports to enhance military training and decision-making. Additionally, the United States and Australia are leveraging generative AI for strategic advantage in the Indo-Pacific, focusing on applications such as enhancing military decision-making, processing sonar data, and augmenting operations across vast distances \citep{Bajraktari2024}. These efforts are accelerating data processing, improving the identification and response to threats, and helping maintain technological and operational superiority by investing in AI talent and reskilling military personnel. These examples demonstrate the transformative potential of LLMs on modern military strategy and operations. 

Besides the US DoD and its allies, the nation's strategic competitors (e.g., China, Russia, North Korea, and Iran) are also exploring the national security applications of LLMs.  For example, China employs Baidu's Erni Bot, an LLM similar to ChatGPT, to predict human behavior on the battlefield to enhance combat simulations and decision-making \citep{McFadden2024}. Some reports indicate Baidu's Ernie Bot surpasses ChatGPT in accuracy and real-time information processing but struggles with political inquiries due to restrictions \citep{Tamin2024}. Additionally, the CopyCat network, suspected to be aligned with the Russian government, leverages LLMs to manipulate media content for disinformation \citep{InsiktGroup2024, InfoSecurity2024}. The CopyCat network plagiarizes, translates, and edits content from legitimate media outlets to spread biased political messages aligned with Russian interests and has been linked to generating sophisticated narratives on the Ukraine conflict and US politics that are particularly difficult for public officials to counteract. Moreover, suspected state-backed hackers from China, Iran, North Korea, and Russia are accused of experimenting with LLMs to assist in cyber operations, e.g., by generating malicious code and content for phishing campaigns \citep{Groll2024}. For instance, North Korea's Kimsuky group uses LLMs to advance their military cyber capabilities by generating content for phishing campaigns, targeting organizations focused on North Korean defense and nuclear issues, thereby enhancing their ability to gather intelligence and exploit vulnerabilities \citep{Reddy2024}. The widespread exploration of LLMs in such capacities highlights the threat the technology poses, particularly in cyber and information warfare, and necessitates the development of countermeasures to safeguard national security.  

\subsection{Relation to Scholarly Perspectives}

Strictly speaking, LLMs are trained to predict the next word in a phrase conditioned upon previously observed words. However, the surveyed use of LLMs for national security applications is not concerned with this task in and of itself. Instead, national security organizations are often concerned with the \textit{emergent} abilities of LLMs. The existence of such emergent LLM abilities as generally conceived is not universally accepted \citep[e.g., see][]{lu2023emergent, schaeffer2024emergent}. Nevertheless, it is undeniable that most national security applications using LLMs seek to extend the technology beyond the task for which the LLMs were trained. 

Natural language tasks are heavily featured in the current efforts of national security organizations, and, in this regard, the capabilities of LLMs are well-established. Namely, \citet{min2023recent} survey the state-of-the-art performance of pre-trained language models on multiple natural language tasks. This survey was recently updated by \citet{minaee2024large} and, given the rapid pace of development, another survey will likely be warranted in the near future. Conspicuously relevant in the defense setting are the capabilities of LLMs for automatic summarization. \citet{zhang2024benchmarking} study LLM performance on news article summarization, finding that a variety of LLMs performed on par with their human counterparts. The authors also determined that prompting and instruction tuning, not model size, dictated model performance in zero-shot summarization. This result presages the general importance of prompt engineering when using LLMs in defense settings. Moreover, given the immense scale of national security operations, it is important to understand how well LLMs summarize longer documents or corpora; these documents often exceed the context window size of standard LLMs. \citet{chang2023booookscore} construct and validate a metric that evaluates LLM performance on such documents, and initial results suggest they perform similarly to humans on the task. \citet{SEI2024} specifically study LLM summarization performance in the DoD acquisition setting with favorable results. 

Alternatively, the self-evident efficacy of LLMs for text generation also implies their efficacy for information warfare. \citet{goldstein2023generative} and \citet{low2023automated} explore how LLMs allow for the automation of influence operations, circumventing the need for labor-intensive \textit{troll farms}, once a feature of the space \citep{woollacott2020}. Coupled with the disinformation threat of other generative AI models \citep{xu2023combating}, the need for an effective detection mechanism for LLM-generated propaganda is imminent. \citet{wan2024dell} propose a prototypical pipeline for doing so. However, the work of \citet{chen2023can} suggest that such messages may be harder to detect than their man-made counterparts, a task of considerable difficulty in and of itself \citep{zhang2019detecting}. \citet{uchendu2023does} also discuss the identification of deepfake texts, concluding that this difficult task may be facilitated via human collaboration. 

Interestingly, the desire to leverage AI for strategy development predates the development of LLMs. \citet{bazin2017} contrived the concept of a virtual Clausewitz, a cognitive computing system with access to the whole of human military thought, that would advise senior military leaders. The development of LLMs has since increased the relevance of such ideas. Notably, COA-GPT \citep{goecks2024coa} is an LLM-based tool designed to automate the COA development phase of the joint planning process \citep{JP5-0}, and the authors test it against baseline reinforcement learning methods in StarCraft II with favorable results. \citet{goecks2023disasterresponsegpt} provide a similar planning tool focused specifically on disaster-relief operations. However,  it is worth noting that both \citet{simmons2024ai} and \citet{Hunter2022} warn against such systems, the former arguing against LLM-recommender systems and the latter against AI more generally in military command decisions. \citet{simmons2024ai} cites the potential for LLMs to escalate and the intractability of their recommendations. \citet{Hunter2022} assert that reliance on AI for command decisions may lower the standards and practice of strategy development. Recent empirical research in tabletop war games corroborates some of these concerns. \citet{rivera2024escalation} perform wargaming experiments whereby each nation state is modeled with a commercial LLM; GPT-4, GPT-3.5, Claude 2, Llama-2 (70B) Chat and GPT-4-Base are tested. Therein, the authors find that the LLMs tend to develop arms-race dynamics, leading to greater conflict and, in rare cases, the use of nuclear weapons. \citet{lamparth2024human} similarly compare how LLMs and humans conduct themselves in a US-China wargame. The authors find that, while the groups behaved similarly, the LLMs were more aggressive. Additionally, \citet{lamparth2024human} found that prompt verbosity affected LLM behavior. Such findings are highly relevant to the future of automated wargaming \citep[e.g., see the \textit{Snow Globe} system of][]{hogan2024open}.

It may be argued that these more ambitious efforts rely upon an LLM having some form of innate knowledge or, at the least, the ability to reflect human logical reasoning. Diverse perspectives exist in the literature on both of these topics. \citet{yildirim2024task} discuss the nature of LLM knowledge, juxtaposing LLM instrumental knowledge with the worldly knowledge of humans. \citet{saba2023stochastic} emphasizes that LLMs cannot be relied upon for factual information because, as currently trained, factual and non-factual information is not differentiated. In this light, if an LLM cannot distinguish fact from fiction, the nature of its knowledge is in question. Moreover, \citet{huang2022towards} consider if and how LLMs execute logical reasoning, concluding that, for LLMs on the scale of GPT-3 175B or higher, reasoning is an emergent behavior, despite their difficulties with complex logical tasks. Alternatively, \citet{ellis2024human} argues that efficient human learning is intrinsically linked to their ability to reason inductively. He claims that out-of-the-box LLMs are deficient in this task but that this shortcoming can be addressed by layering a Bayesian model on top of a pre-trained LLM. Setting aside the actual nature of LLM knowledge and reasoning, many researchers and practitioners seek to leverage LLMs in various game-playing and strategic decision-making tasks. \citet{xu2024survey} and \citet{chen2024large} provide comprehensive surveys, respectively, that provide further context on LLM use for national security planning. 

Finally, in view of the momentous effects associated with national security operations, we would be remiss if we excluded interpretable, explainable, or adversarial machine learning from this discussion. Each of these sub-disciplines is consequential to using LLMs for strategic decision-making. Both interpretable and explainable machine learning focus on tractability but through distinct means. Interpretable ML seeks to build simple yet powerful models, and explainable ML endeavors to identify simple yet faithful approximations of a black-box model. To date, LLMs are best viewed as a black-box model, implying that explainable techniques \citep[e.g., see the Deep SHAP values of][]{lundberg2017unified} are most readily applicable; however, the development of interpretable image classifiers \citep[e.g., see][]{chen2019looks} suggest that the development of high-performing, interpretable language models may be feasible. Applying either approach is useful if it can successfully identify undesirable behavior, e.g., bias in automatic summarization \citep{zhou2023characterizing}. Alternatively, adversarial ML focuses on the corruption and defense of ML methods in the presence of an opponent. In the context of LLMs, training-time attacks can be used to modify output in accordance with the attacker's objectives. \citet{bagdasaryan2022spinning} provide a concrete example whereby poisoning attacks to the training data encourage LLMs to spin automatically generated summaries according to the adversary's point of view. Deployment-time attacks against a pre-trained LLM focus on modifying input data to affect a desired output. \citet{raina2024llm} explore such attacks against assessment LLMs whereby a short universal phrase is appended to a phrase to disrupt model performance. The threat of such attacks to national security is self-evident. If an attacker can manipulate an LLM that informs command decision-making, traps may be laid in advance so that their adversary acts per their desires. For additional information on adversarial ML, we point the interested reader to \citet{oprea2023adversarial}.

\section{Discussion} \label{sec3}

National security professionals are generally proceeding cautiously with LLM-based technology. Although some seek to use LLMs as a foundation upon which to build a virtual Clausewitz, the primary focus of national security organizations is to use LLMs for natural language tasks, a function in which their capabilities are well-established. This calculated approach is enshrined in the strategic documents and international agreements adopted by many countries. For example, the \citet{DoD_RAI_Strategy} set forth a framework for AI implementation conditioned upon its responsible, equitable, traceable, reliable, and governable use. Similar tenets are foundational to the approach toward emerging and disruptive technologies set forth by \citet{NATO2022}. The Political Declaration on Responsible Military Use of Artificial Intelligence and Autonomy, endorsed by 54 countries, also echoes analogous themes \citep{State2024}. In our estimation, such deliberate action is appropriate given the nascent nature of LLMs and our incipient understanding of their performance.

Utilizing LLMs for standard natural language tasks, e.g., automatic summarization, is admittedly less intriguing than their use for generating automated campaign plans. However, we contend that if these less alluring functions were embedded across the national security ecosystem, the effects would be transformational. Defense organizations are characterized by massive bureaucracies. In these bureaucracies, many offices are tasked with aggregating information and distilling insights to send to the next level in the bureaucratic hierarchy. Others are responsible for aggregating information, proposing courses of action for a commander's decision, and disseminating these decisions via written documentation. When conducted exclusively by humans, these tasks are incredibly slow and cumbersome. Despite the adoption of information-age technology, many defense organizations have struggled to effectively manage data \citep{DoD2023}, forcing staffs to rely on manual inquiry and collection. Moreover, many national security operations, particularly those below armed conflict, inhibit the development and use of structured databases. National objectives are often qualitatively defined without a clear quantitative analog \citep[e.g., see the Hamlet Evaluation System discussed by][]{mushen2014we}, allocated resources may rapidly change due to shifting national priorities, and personnel are transient (e.g., service-member deployments). 

Much of the data available to national security organizations is thus unstructured, often in the form of text or images. Current practices require a human to ingest, summarize, and distill insight from this information, thereby limiting decision-making to the speed of human comprehension. Therefore, LLM-based automatic summarization may be particularly well-suited for national security applications. Such models can ingest and summarize a significant amount of unstructured information much quicker than their human counterparts. Alternatively, in use cases where structured databases exist, LLMs may be used as a component within broader systems that translate text inquiries into code, thereby allowing a layman to readily interact with data. For example, a non-technical analyst may enter a text query that triggers the LLM to generate structured query language (SQL) code for data set creation, R code for visualization, and text suggestions for more follow-up statistical analyses. 

Integrating LLMs with alternative probabilistic, statistical, and ML methods can further streamline information processing and analysis by breaking down queries into manageable steps and leveraging external tools for precise calculations. For instance, combining LLMs with statistical forecasting methods can enhance the accuracy of intelligence predictions, and using LLMs alongside supervised ML techniques can improve data classification and analysis. Moreover, through interaction with an LLM that has access to the relevant information, an analyst can conversationally explore a compendium of structured and unstructured data to derive novel insights. By facilitating knowledge-based work, LLMs may allow commanders to reduce their bureaucratic footprint and reallocate personnel from administrative to operational roles. LLMs may also enable the deconstruction of information silos fashioned within the bureaucracy due to limited human awareness and communication. Coupling these potentialities with the speed of LLM processing, national security organizations may be able to rapidly accelerate and improve their decision-making cycles; \citet{BusinessInsider2023} provides an example in USAF operations wherein an LLM needed 10 minutes to complete a task that typically requires days of man-hours. 

Nevertheless, hallucinations remain a potential threat, and this risk must be appropriately mitigated. Recent research by \citet{nahar2024fakes} suggests that simply warning users about hallucinations improves their ability to detect them. Future academic studies should also compare LLM-induced hallucinations to errors induced by transmission chaining in large bureaucracies; such studies would provide a baseline to determine the degree to which LLM hallucinations deviate from errancy in the status quo. Alternative comparative studies could view the problem from a decision-theoretic perspective, investigating how well each method summarizes data and how this affects decision-making. LLMs may fall short of expectations, or their utilization in national security staffs may evoke unforeseen consequences; however, they hold great promise in expediting sluggish bureaucratic processes.

Conversely, the direct use of LLMs for real-world, strategic-level planning is more fraught. Recently developed LLMs have shown ``sparks" of artificial general intelligence (AGI) \citep{bubeck2023sparks}, obtaining near human performance on tasks from a variety of fields (e.g., medicine, law, and psychology); however, we contend that national security planning is a function of a different character. No textbook contains the right answers from which an LLM can learn. The work of military historians, along with the treatises of Sun Tzu, Jomini, and Clausewitz, contain valuable information that may allow an LLM to craft a written strategic document comparable to the status quo. However, such documents are aspirational and are useful insofar as the humans executing the strategy achieve unified action across domains throughout the strategic, operational, and tactical levels of command. Achievement of this goal has eluded even the most skilled and experienced national security leaders. Thus, we are skeptical that a recipe for its attainment is embedded even within even the whole of written human text. That is not to say that AI cannot attain superhuman abilities for the task or that LLMs are not the building blocks to achieve it, but we believe that more than sparks of AGI would be required. Alternatively, we are more optimistic about utilizing contemporary LLMs to train national security strategists and planners. Their use in wargaming allows for individual study and personalized instruction heretofore lacking in professional military education. Automated, qualitative wargames \citep[e.g., see][]{hogan2024open} may be calibrated for specific scenarios and adversaries, transcripts may be reviewed later with instructors to facilitate learning, and students may replicate the same wargame autonomously to rectify their mistakes. We find this last feature of replication particularly compelling. Military planners often struggle to cope with the vast degree of uncertainty inherent in their operations and do not understand the distinction between decision and outcome quality. If properly executed during wargame training, LLMs may be able to partially rectify these issues. 

Based on the widespread efforts by national security organizations, we believe LLMs will further the AI-induced change in the character of conflict. In particular, LLMs will disrupt the status quo under the threshold of armed conflict. This is already being seen through their use in the development of nation-state-sponsored propaganda. Disinformation campaigns are becoming cheaper at scale, and automated propaganda is harder to discern. This trend will only continue as LLMs are paired with alternative generative AI methods (e.g., those creating images and videos). Moreover, as LLMs are incorporated into national security processes, they will further expand the relevance of adversarial machine learning and instantiate a new cyber-warfare arena. The black-box nature of existent LLMs increases this threat, making the incorporation of interpretable and/or explainable models imperative to enabling attack-and-vulnerability detection. Given the breadth of contemporary LLM initiatives and applications, the aforementioned threats and mitigation efforts are relevant across the competition continuum.

\section{Conclusion} \label{sec4}
Integrating LLMs into national security operations presents unprecedented opportunities and significant challenges. As evidenced by their varied applications across the DoD and other world defense organizations, LLMs have the potential to revolutionize the efficiency (and effectiveness) of national security operations. The potential benefits are substantial. LLMs can automate and accelerate information processing, enhance decision-making through advanced data analysis, and reduce bureaucratic inefficiencies. Their automatic summarization capabilities can streamline the creation of operational documents and reports, while their integration with probabilistic, statistical, and ML methods can improve accuracy and reliability, e.g., combining LLMs with Bayesian techniques can provide more robust threat predictions. Conversely, deploying LLMs into national security organizations does not come without risks. More specifically, the potential for hallucinations, the ensuring of data privacy, and the safeguarding of LLMs against adversarial attacks are significant concerns that must be addressed. These risks are particularly concerning in high-stakes decision-making environments where the accuracy and integrity of information are crucial (e.g., armed conflict). Addressing these challenges is critical to protecting sensitive information, preventing malicious exploitation, and avoiding potential national security catastrophes.

The broader implications of LLM integration into national security organizations are profound. Beyond immediate defense applications, LLMs have the potential to influence strategic planning, international relations, and the broader geopolitical landscape. The purported ability of nations to leverage LLMs for disinformation campaigns and cyber operations emphasizes the need to develop appropriate countermeasures and continuously scrutinize and update AI security protocols. While LLMs have demonstrated ``sparks” of AGI, we contend that their current capabilities are best suited for supporting roles rather than leading strategic decisions. Their use in training and wargaming can provide valuable insights and personalized learning experiences for military personnel, helping to bridge knowledge gaps and improve operational readiness.

A cautious and calculated approach must be taken by national security professionals with regard to integrating LLMs into their operations. Deliberate integration, guided by frameworks for responsible AI use, is both appropriate and necessary for harnessing the potential benefits of LLMs while mitigating associated risks. Moving forward, continued research and collaboration between defense, academic, and commercial entities is essential to fully realize the benefits of LLMs. National security professionals must ensure that, while they explore how these powerful tools can be leveraged to enhance national security capabilities, they simultaneously focus on safeguarding against potential threats. This balanced approach will enable them to navigate the complexities of AI integration and establish strategic advantage in an increasingly contested and technologically advanced world.



\section*{Disclaimer}
The views expressed in this article are those of the authors and do not reflect the official policy or position of the United States Air Force, United States Department of Defense, or United States Government. This work is approved for public release (distribution unlimited) in accordance with PA\# USAFA-DF-2024-514.

\bibliography{biblio.bib}

\end{document}